# Strong Light Confinement in Rod-Connected Diamond Photonic Crystals

MIKE P. C. TAVERNE[1,2] YING-LUNG D. HO[1,*] XU ZHENG,[1] LIFENG CHEN,[1] CHEN-HSUAN N. FANG,[1] JOHN RARITY[1,3]

[1]Department of Electrical and Electronic Engineering, University of Bristol, Merchant Venturers Building, Woodland Road, Bristol, England, UK
[2]e-mail: Mike.Taverne@bristol.ac.uk
[3]e-mail: John.Rarity@bristol.ac.uk
*Corresponding author: Daniel.Ho@bristol.ac.uk

**We show that it is possible to confine light in a volume of order $10^{-3}$ cubic wavelengths using only dielectric material. Low-index (air) cavities are simulated in high index rod-connected diamond (RCD) photonic crystals. These cavities show long storage times ($Q$-factors $>10^6$) even at the lowest volumes. Fabrication of such structures could open up a new field of photon level interactions.** © 2018 Optical Society of America

The interaction of light with nonlinear materials and light emitters can be enhanced by confining the light in a region of space small compared to the cubic wavelength [1]. Although this can be achieved by placing the emitter close to metallic structures supporting plasmons [2], it necessarily incurs a loss penalty. If equivalent confinement could be achieved in a loss free dielectric environment new high efficiency single photon sources [3], spin photon interfaces [4] and single emitter strong coupling at elevated temperatures [5] might be achieved. Here, we investigate the limits to cavity volume when light confinement is provided by a 3D photonic crystal showing a full photonic bandgap.

In contrast to previous work [6] where we looked at cavities formed by *adding* high refractive index material to form point defect cavities, here we *remove* material to create low-index cavities. This results in much smaller effective cavity volumes ($V_{eff}$) and much higher field enhancement effects. We study rod-connected diamond (RCD) 3D photonic crystals (PhCs) [6-9] and incorporate a spherical low-index cavity ($n_{def}$ = 1) to show ultra-small mode volumes (~ $10^{-3}(\lambda/n_{def})^3$). This low-index cavity mode volume is much smaller than mode volumes obtained for high index defects in inverse RCD structures (~0.06 $(\lambda/n_{def})^3$ [6], and face-centered-cubic (FCC) woodpile cavities (~0.1$(\lambda/n_{def})^3$ [10]. Furthermore it is more than one order of magnitude smaller than the mode volume obtained in direct RCD (~0.09 $(\lambda/n_{def})^3$ [7]), using cylindrical low-index defects ($n_{def}$ = 1).

Recently, 3D photonic crystal structures with complete photonic bandgaps (PBGs) have been fabricated using direct laser writing, exploiting two-photon polymerization (2PP) based 3D lithography [11]. The technique is ideal for fabricating non-layer based structures such as RCD, and could allow selective writing of defects containing fluorescent material at the antinodes of cavities.

RCD crystals exhibit the largest PBG known to date [12] and here we a focus on inverse RCD [6] structures as they are more manufacturable. The cubic unit-cell (of size $a_u$) is illustrated in Fig. 1 showing the relation to the Brillouin zone geometry and the defect position. We have studied high refractive-index materials [e.g. gallium phosphide GaP] in a low index (Air) background material showing a dielectric contrast of 3.3:1. This results in an optimized normalised air-rod radius ($r/a_u$= 0.26) and full photonic bandgap that has a relative frequency width $\Delta\omega/\omega_0$ ~25% with a midgap normalised frequency $a_u/\lambda \sim 0.59$.

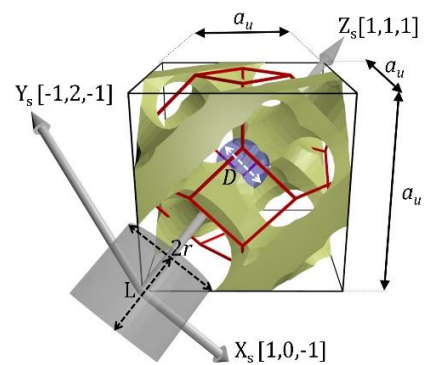

Fig. 1. The cubic unit cell (size $a_u$) of an inverse rod-connected diamond photonic crystal with high refractive index region shown in grey-green. The spherical air defect is shown as a transparent blue sphere of diameter $D = 2r_d$. The first Brillouin zone of the FCC lattice (red) is also shown. Lattice vectors corresponding to our simulation axes ($X_s$, $Y_s$, $Z_s$) are shown as grey arrows and a light grey cylinder illustrates a rod in the original RCD lattice.

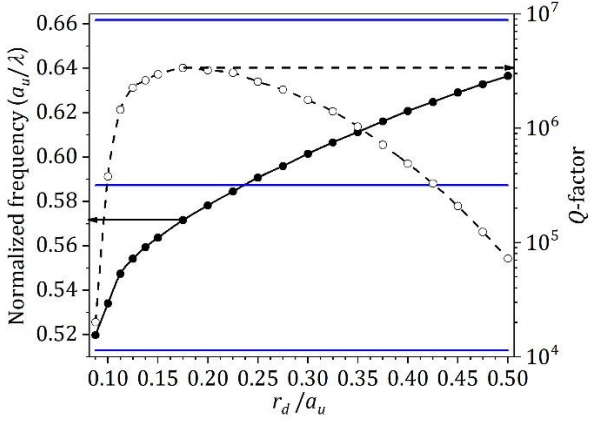

Fig. 2. The normalized frequency of resonance peaks ($a_u/\lambda$) (solid circle and line) and Q-factors (open circle, and dashed line) obtained for the various normalized radii ($r_d/a_u$) assuming 3.3:1 index contrast (eg air sphere defects in GaP). The blue lines indicate the limits to the full bandgap ($a_u/\lambda \sim 0.513$–$0.66$) and midgap frequency $a_u/\lambda \sim 0.59$.

The optimal radius of the air-rods $r$ is about 2.4 times larger than in the direct RCD case making manufacture more feasible [6]. A spherical air cavity is then introduced in one leg of the central unit cell of a 10×10×10 unit cell simulation volume.

The simulations were conducted using the finite-difference time domain (FDTD) method [13] with a variety of low-index defect sizes. After an impulse excitation in the defect, the decay of the electric field amplitude is monitored as a function of time. The Q-factors (the ratio of resonant wavelength to line width $Q = \lambda_{res}/\Delta\lambda$) can then be estimated by analyzing the field decay in the frequency domain via the fast Fourier transform (FFT). We explored excitation dipoles aligned to the three axes ($X_s, Y_s, Z_s$) but see the best results from the $Z_s$ oriented dipole as mode confinement is best in the $X_s$ - $Y_s$ plane.

Any perturbation disrupting the translational symmetry can act as a defect. Hence, precise tuning of the cavity resonances can be achieved by varying the amount of perturbation through the modification of defect sizes in 3D PhCs. Here, we varied the air sphere defect radius $r_d$ from $0.0875a_u$ to $0.15a_u$ in $0.0125a_u$ steps and $0.15a_u$ to $0.5a_u$ in $0.025a_u$ steps. As shown in Fig. 2, the normalized resonance frequency ($a_u/\lambda$) of the defect cavities increases with defect size. The corresponding Q-factors obtained for the different defect sizes are shown in Fig. 2. The defect radius, $r_d = 0.175a_u$, shows a maximum Q-factor ($Q \sim 3.4 \times 10^6$) when the resonance frequency is close to the centre of the bandgap. When the resonance frequency is close to the band edge the Q-factors are reduced. For example with a defect radius $r_d = 0.0875a_u$, Q-factors are reduced to $Q \sim 2 \times 10^4$. Hence, the appropriate selection of the size of defects is critical in optimizing the confinement of the electric field within the cavities.

Having determined the resonant frequency, a cavity mode on resonance can be visualized using a series of single frequency snapshots. For an air sphere defect, the confinement of the normalized square of the electric field ($|E|^2$) distribution are illustrated in Fig. 3 for electric field dipoles oriented along the $Z_s$ directions [field direction arrow visible in Figs. 3 (a), 3(b), 3(f), and 3(g)]. The regions where $|E|^2 \geq 0.40(|E|^2)_{max}$ are shown with respect to the lattice in Fig. 3(a) for a defect size $r_d = 0.1a_u$ and Fig. 3(f) for a defect size $r_d = 0.175a_u$. 2D cross-sections of the the normalized square of the electrical field ($|E|^2/(|E|^2)_{max}$) taken along the $X_s$, $Y_s$ and $Z_s$ planes going through the electric field maximum are shown in Figs. 3(b)–3(e) for $r_d = 0.1a_u$ and Figs. 3(g)–3(j) for $r_d = 0.175a_u$.

For the close up cross-sections, we see the essentially digital nature of our surfaces. We used a graded mesh with a high resolution near the defect (smallest cell size: $0.0125a_u$) and tapering off to a lower resolution further away from it (largest cell size: $0.0683a_u$).

As the defect radius increases the truncated shape of the cavity varies and the electric field mode size increases. The electric field maxima tend to form rings in the $Z_s$ plane along the edges of the sphere defect (*defect edge rings*). We see that for the smallest sphere cavity, when the *defect edge rings* are close together, the electric field maxima form a single ring close to the centre of the cavity [Figs. 3(a), and 3(b)] illustrating similar behaviour to slot waveguides. However, for larger cavities, when the *defect edge rings* are further apart, the field maxima tend to form two separate rings localized near the defect edges [Figs. 3(f), and 3(g)].

Observation of modified spontaneous emission and nonlinear optical effects in microcavities depend on their high-quality factors ($Q$) and small mode volumes ($V_{eff}$), that is, with a high ratio of $Q/V_{eff}$. However, there is a point where reducing the size of defect cavities leads to a poorer confinement of the field and lower $Q$ and thus an maximum $Q/V_{eff}$ point can be found. Hence, we calculate the mode volume as a function of defect cavity radius using the following definition of the effective mode volume ($V_{eff}$) and normalized mode volume ($f_{opt}$):

$$V_{eff} = \frac{\iiint \varepsilon(r)|E(r)|^2 d^3r}{\varepsilon(r_{max})[|E(r)|^2]_{max}} \quad (1)$$

$$f_{opt} = \frac{V_{eff}}{[\lambda/n(r_{max})]^3} \quad (2)$$

where $r_{max}$ is the position of the maximum electric field amplitude. Note that this definition is based on [14] and is different from the more commonly used definition found in [15]. It is a more generalized formula, which takes into account cases where the maximum electric field (relevant for photon-atom coupling calculations) is found in a low index region, despite the maximum electric energy density potentially being in a high index region. For the low index case normalized mode volume can be minimized by increasing the mode maximum electric field and localizing the mode maximum in the low index region.

The cavity volumes calculated using Eq. (1) and Eq. (2), are shown in Fig. 4. In general, the mode volume decreases with decreasing defect size, as expected, except for a small reverse at $r_d = 0.225a_u$ (possibly arising from the defect sphere's diameter exceeding the length of the RCD rods and therefore passing the centre of the inverse tetrahedral structures) and, also, for the smallest defect size ($r_d = 0.0875a_u$) which sits entirely within the rod (ie is not manufacturable). The smallest volume is achieved with a defect radius $r_d = 0.1a_u$, with $V_{eff} \sim 0.0035(\lambda/n)^3$, although we see that for this defect the Q-factor has dropped to $\sim 3.8 \times 10^5$. A better $Q/V_{eff}$ is achieved at $r_d = 0.1125a_u$ where $V_{eff} \sim 0.0036(\lambda/n)^3$ can be obtained, while still retaining $Q \sim 1.5 \times 10^6$. However, this is a full bandgap material. Hence, the Q-factor we measured is essentially limited by the number of periods in our simulation [6, 10, 16], and could be further increased while maintaining a small mode volume $V_{eff}$. Of course in all fabricated structures the limit to Q-factor would likely be due to material absorption loss.

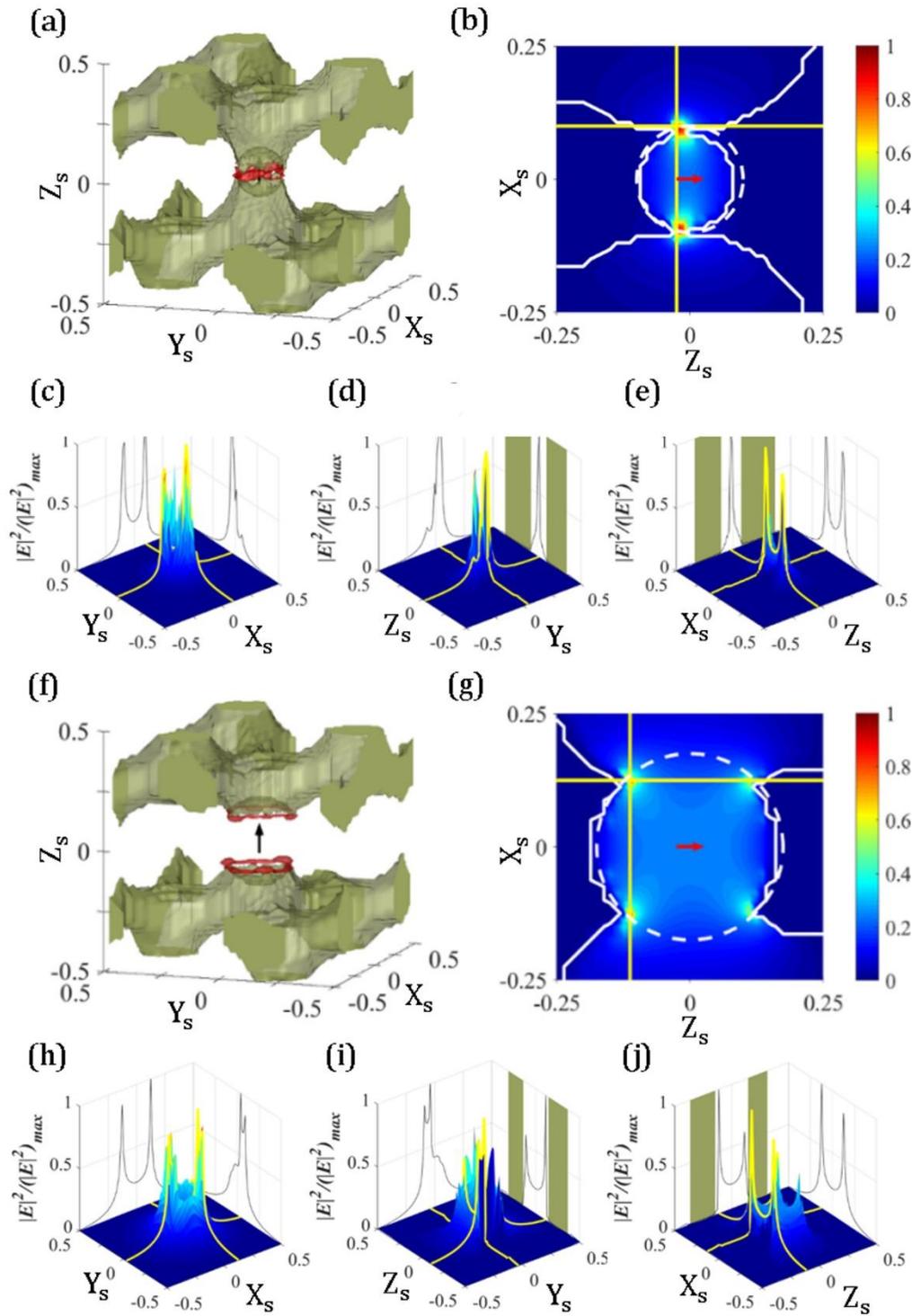

Fig. 3. (a, f) Isosurfaces of the high index material (grey-green, $n$=3.3) and normalized square of the electric field distribution (red regions corresponding to $|E|^2/(|E|^2)_{max} \geq 0.40$) with low-index spherical cavities (voids, $n_{def}$=1) corresponding to defect sizes $r_d = 0.1a_u$ and $0.175a_u$, respectively. The cavities were excited by a dipole with an electric field oriented along the $Z_s$ direction (cf. black and red arrows in (a),(b),(f),(g)), using a short Gaussian pulse. (b-e), (g-j): 2D cross-section plots of the normalized square of the electric field ($|E|^2/(|E|^2)_{max}$) done along the $X_s$, $Y_s$, $Z_s$ planes and going through the electric field maxima of each field distribution. In (c-e) and (h-j), an additional 1D cross-section is plotted on the sides with the green bands highlighting the high-index regions along the used cross-section lines (drawn in yellow on top of the 2D surfaces). The white dashed circles in (b), (g) show the location of the air sphere defect. The yellow solid lines in (b-e), (g-j) show the location of the maximum electric field. (b) and (g) are a view from the top of (e) and (j) respectively, with thick white lines representing the contour lines of the material distribution.

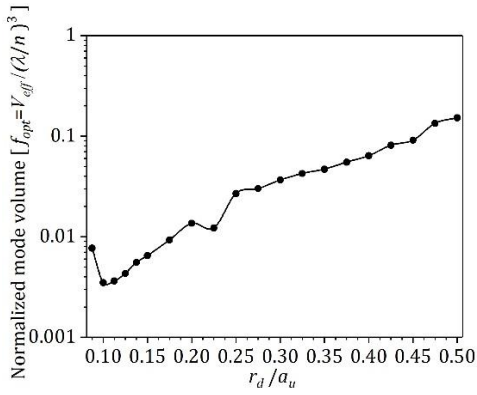

Fig. 4. The normalized mode volumes obtained for the various normalized radii ($r_d/a_u$) and refractive index values ($n_{def}$ = 1.0) of the sphere defects.

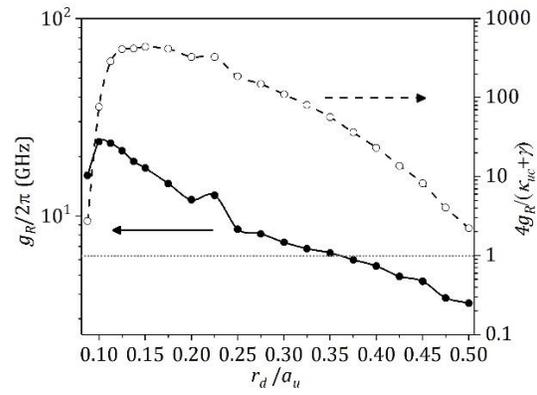

Fig. 5. $g_R$ coupling rate (solid circle and line) and strong/weak-coupling criteria $4g_R/(\kappa_{uc}+\gamma)$ (open circle, and dashed line) of the studied inverse RCD defects plotted against the normalized radius ($r_d/a_u$) of sphere defects for a diamond NV -center.

We can now estimate the possibility of achieving strong coupling using typical two level emitters. The coupling strength $g_R$ of the cavity mode with a quantum emitter placed inside the defect can be calculated using the following formula [6]

$$g_R = \frac{3c_0 \gamma_{os} n_{def}}{8\pi f_{opt} \lambda_{os} n_{os}} \quad (3)$$

where $c_0$ is the speed of light in vacuum, $\gamma_{os}$ and $\lambda_{os}$ are the spontaneous emission rate and wavelength of the considered quantum emitter and $n_{os}$ is the refractive index directly surrounding the emitter. The weak- and strong-coupling criteria $4g_R/(\kappa_{uc} + \gamma_{os})$, where $\kappa_{uc} = 2\pi c_0/(\lambda_{os} Q)$ is the decay rate of the uncoupled cavity, can also be calculated. Here, we consider coupling to the zero phonon line (ZPL) of diamond NV -centers, for which $n_{os}$ = 2.4, $\lambda_{os}$ = 637 nm and $\gamma_{os} = \gamma_{ZPL} \sim 4\% \cdot \gamma_{total} = 3.3 \times 10^6$ rad/s because emission into the ZPL constitutes only 4% of the total spontaneous emission [10, 16]. Figure 5 shows the $g_R/2\pi$ = 16, 24, 23, and 15 GHz in defect size of $r_d$ = $0.0875a_u$, $0.1a_u$, $0.1125a_u$, and $0.175a_u$, respectively; all considered size of defects have $4g_R/(\kappa_{uc} + \gamma_{os})> 1$ and should therefore lead to strong coupling.

In summary, we have studied inverse rod connected diamond photonic crystals (RCD PhCs) formed in high-index-contrast materials [3.3:1.0 (GaP-air)]. The inverse structures appear more easily fabricated while still retaining wide bandgaps. Introducing air sphere defect cavities at optimized locations within the crystal we can create high-$Q$ cavities with ultra-low mode volumes ($V_{eff}$). Using FDTD simulations, we find that an air sphere defect, $r_d$ = $0.1a_u$ gives the best result with a mode volume $V_{eff} \sim 0.0035(\lambda_{res}/n_{air})^3$ and $Q \sim 3.8\times10^5$, which corresponds to a $Q/V_{eff}$ ratio $\sim 1.1 \times 10^8(\lambda_{res}/n_{air})^{-3}$ with a normalised resonance at $a_u/\lambda \sim 0.534$. In our full set of results, we have also simulated cavities with up to 3.6:1.0 index-contrast and see mode volumes down to $V_{eff} \sim 0.0025(\lambda_{res}/n_{air})^3$ with higher $Q$-factors up to $\sim 1.4\times10^7$. To our knowledge, this is a record low mode volume for defect cavities in 3D photonic crystal structures.

Such a high-$Q$ cavity with ultra-small mode volume could be used to demonstrate strong coupling at elevated temperatures [5], while coupled cavities of this type could form broad bandwidth, lossless, wavelength scale optical circuits in a fully 3D photonic crystal microchip. Additionally, these microchips could allow the development of ultrasensitive gas sensors, guiding and confining light in air or low refractive index material, and may also find applications in solar energy trapping and harvesting [17].

**Funding.** EPSRC grants EP/M009033/1 and EP/M024458/1.

**Acknowledgment.** This work was carried out using the computational facilities of the Advanced Computing Research Centre, University of Bristol - http://www.bris.ac.uk/acrc/.